\documentstyle[pre,aps,multicol,epsf]{revtex}
\begin{document}
\draft
\title{The Elastic Behavior of Entropic ``Fisherman's Net''}
\author{Oded Farago and Yacov Kantor}
\address{School of Physics and Astronomy, Tel Aviv University, Tel
  Aviv 69 978, Israel}
\maketitle

\begin{abstract}
A new formalism is used for a Monte Carlo determination of the elastic
constants of a two-dimensional net of fixed connectivity. The net is
composed of point-like atoms each of which is tethered to six
neighbors by a bond limiting the distance between them to a certain
maximal separation, but having zero energy at all smaller lengths. We
measure the elastic constants for many values of the ratio $\gamma$
between the maximal and actual extensions of the net. When the net is
very stretched  ($\gamma\sim1$), a simple transformation maps the
system into a triangular hard disks solid, and we show that the
elastic properties of both systems, coincide. We also show that the
crossover to a Gaussian elastic behavior, expected for the
non-stressed net, occurs when the net is more loose ($\gamma\sim 3$).

\end{abstract}
\begin{multicols}{2}
\narrowtext

Materials like rubber and gels are formed when polymers or monomers
are cross-linked into macroscopically large networks. Due to the small
energetic differences (of the order of $kT$) between the allowed
microscopic configurations of these materials, their physics is
primarily determined by entropy, rather than energy. This has been
recognized long ago, and the peculiar physical properties of rubber
and gels, in particular their great flexibility, are attributed to
this microscopic feature. The classical theories of rubber elasticity,
for instance, deal with Gaussian networks in which the internal
elastic energy is completely ignored and the strands between
cross-links are viewed as entropic springs \cite{trelo}. These
theories, however, do not explain well the elastic behavior of
networks of certain types. Perhaps the most known unresolved problem
in this field of research, is the question of the critical elastic
behavior of random systems near connectivity threshold. Most of the
numerical works which aimed to investigate this issue during the last
twenty years, concerned with the {\em energetic}\/
elasticity~\cite{energy}. Recent studies \cite{plis}, however,
suggested that close to the gel-point elasticity is dominated by its
{\em entropic}\/ component. A completely different aspect of entropic
elasticity which has been studied much less, is the behavior of highly
connected networks, well above their connectivity threshold. The
classical theories are inappropriate in this connectivity regime,
since the strands between the junctions are very short and do not
resemble Gaussian springs.      

When the {\em boundary}\/ of a thermodynamic system is {\em
  homogeneously}\/ deformed, the distance between any two boundary
points which prior to the deformation were separated by $\vec{R}$,
becomes   
\begin{equation}
r=[R_iR_j(\delta_{ij}+2\eta_{ij})]^{1/2}, 
\label{trans}
\end{equation}
where the subscripts denote Cartesian coordinates and summation over
repeated indices is implied. The quantities $\eta_{ij}$ are the
components of the {\em Lagrangian strain tensor}, while $\delta_{ij}$
is the Kr\"{o}necker delta. The elastic behavior of the system is
characterized by the {\em stress}\/ tensor, $\sigma_{ij}$, and the
tensor of {\em elastic constants}, $C_{ijkl}$, which are the
coefficients of the free energy density expansion in the strain
variables
\begin{equation}
f(\{\eta\})=f(\{0\})+\sigma_{ij}\eta_{ij} 
+{1\over 2}C_{ijkl}\eta_{ij} \eta_{kl}+\ldots\ .
\label{expan}
\end{equation}
Measuring the elastic constants is much more difficult in
entropy-dominated systems than in energy-dominated ones. In the latter
one needs to calculate energy variations around well defined ground
states. In the former, on the other hand, different microscopic
configurations possess similar energies. Entropy in this case is
essentially the (logarithm of the) number of allowed microscopic
configurations. Measuring the variations of this quantity in response
to external deformations applied on the system, is usually very
complicated. In order to simplify this task, and due to the fact that
the exact energy details are quite irrelevant in entropy-dominated
systems, the inter-atomic interactions in such systems are often
modeled by ``hard'' potentials. Excluded volume effects, for instance,
can be modeled by the hard spheres repulsion, while chemical bonds can
be replaced by inextensible (``tether'') bonds which limit the
distance between the bonded monomers, but have zero energy at all
permitted distances \cite{kkn}. The energy of all the microscopic
configurations in such models, which are called {\em ``athermal''}, is
the same, and their physics, therefore, is exclusively governed by
entropy considerations. It is interesting to note that although
athermal models have been investigated quite extensively in polymer
and soft matter physics, the elastic properties of many of them are
not well understood. Hard spheres systems, for instance, are studied
for already more than 40 years \cite{gastrus}. They were, in fact, the
first systems for which Metropolis et al.\ performed the first Monte
Carlo (MC) simulations in 1953 \cite{metro}. The phase diagram of hard
spheres, which is a function of a single parameter, their volume
fraction, had been fully explored both in simulations and
experiments \cite{space}. Yet, despite of the numerous works dedicated
to {\em elasticity of}\/ hard spheres \cite{elasths}, the accuracy of
the values of their elastic constants still leaves much to be desired.

In the canonical ensemble, the elastic constants can be related to the
mean squared thermal fluctuations of the stress tensor components
(just as the heat capacity is proportional to the mean squared energy
fluctuation). This relation, first expressed by Squire et al.\
\cite{shh}, can be used for a Monte Carlo determination of the
elastic constants. The method is known as the ``fluctuation
method''. The instantaneous stress, measured at a given microscopic
configuration, is associated with the mean force (averaged over the
entire volume) acting on the atoms \cite{lla}. The local forces
originate from external potentials and inter-particle interactions. In
entropy-dominated systems, these forces are usually very small. They
become extremely large only over very short time intervals when atoms
come to the close vicinity of each other or when bonds are
sufficiently stretched. Model with hard potentials can be regarded as
the limiting case in which these time intervals vanish, while at the
same time the instantaneous forces become infinitely large, keeping
the rate of momentum exchange between atoms fixed. It is obvious that
the stress in such systems must be related to the two-point
probability densities of contact between spheres and occurrence of
bond stretching, while the elastic constants (stress fluctuations)
must be related to the corresponding four-point probability
densities. Indeed, we have recently succeeded to formulate the exact
relations. We obtained expressions enabling a direct measurement of
the entropic contribution to the elastic constants, and demonstrated
the accuracy and efficiency of the method using this formalism on 
three-dimensional hard spheres systems \cite{fk}. In this paper we
apply this new formalism to measure, by means of MC simulations, the
stress and elastic constants of topologically simple networks. We
consider a ``toy model'' consisting of a two-dimensional (2D) network
of atoms forming a triangular ``fisherman's net'' (FN): atoms are
point-like, i.e., have no excluded volume, and each one of them is
connected to six neighbors by a ``tether'' limiting the maximal
distance between the atoms, but otherwise not exerting any force. The
FN is a highly connected network, whose physical behavior is
entropy-dominated. Very few studies were devoted to systems of this
type, and it is indeed quite clear that the determination of the
elastic constants of systems similar to the FN, is far from being
trivial.        

The FN is six-fold symmetric when it is equally stretched along all
the spatial directions. Its elastic properties in this reference state
should be as of an isotropic system \cite{ll2}: Its stress tensor is
diagonal with the elements $\sigma=\sigma_{xx}=\sigma_{yy}=-P$, where
$P$ is the {\em negative}\/ external pressure (stretching) one needs
to apply to the boundaries in order to balance the forces exerted by
the net. Only four elastic constants of the net do not vanish:
$C_{11}=C_{xxxx}$, $C_{22}=C_{yyyy}$, $C_{12}=C_{xxyy}=C_{yyxx}$ and
$C_{44}=C_{xyxy}=C_{yxyx}=C_{xyyx}=C_{yxxy}$. Due to the isotropic
nature of the system, only two of them are independent, and they
satisfy the relations: $C_{11}=C_{22}$, and, $2C_{44}=C_{11}-C_{12}$
\cite{wallace}. It is quite common to describe the elastic properties
of isotropic systems in terms of the {\em shear}\/ modulus, $\mu$, and
the {\em bulk}\/ modulus, $\kappa$, defined by: $\mu=C_{44}-P$, and,
$\kappa=\frac{1}{2}(C_{11}+C_{12})$. When these quantities are
positive, the isotropic system is mechanically stable \cite{zhoujoos}.
\begin{figure}[htb]
\epsfysize=14\baselineskip
\centerline{\vbox{
      \epsffile{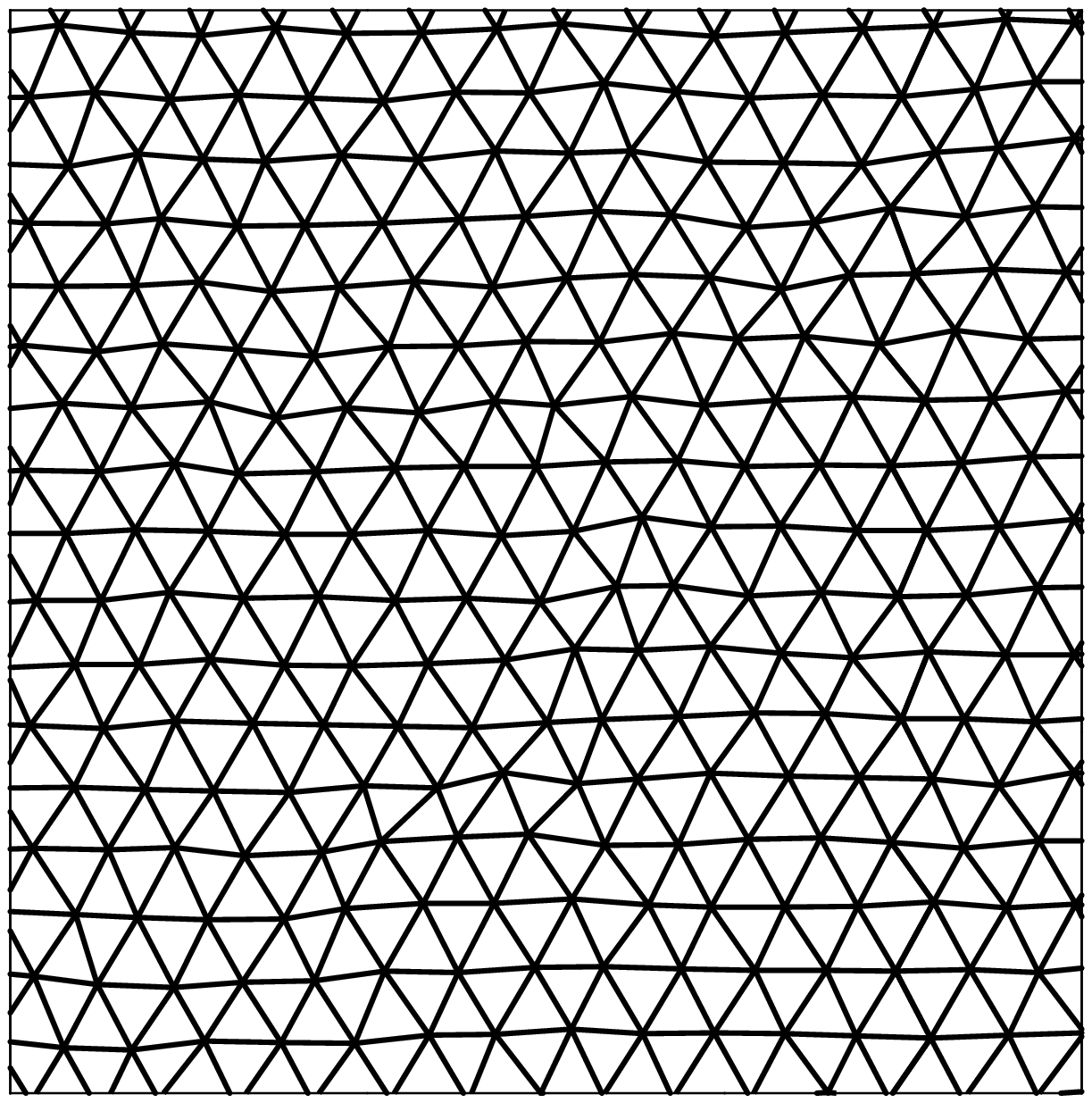}}}
%\vspace{-1.0cm}
\centerline{\large (a)}
\epsfysize=14\baselineskip
\centerline{\vbox{
      \epsffile{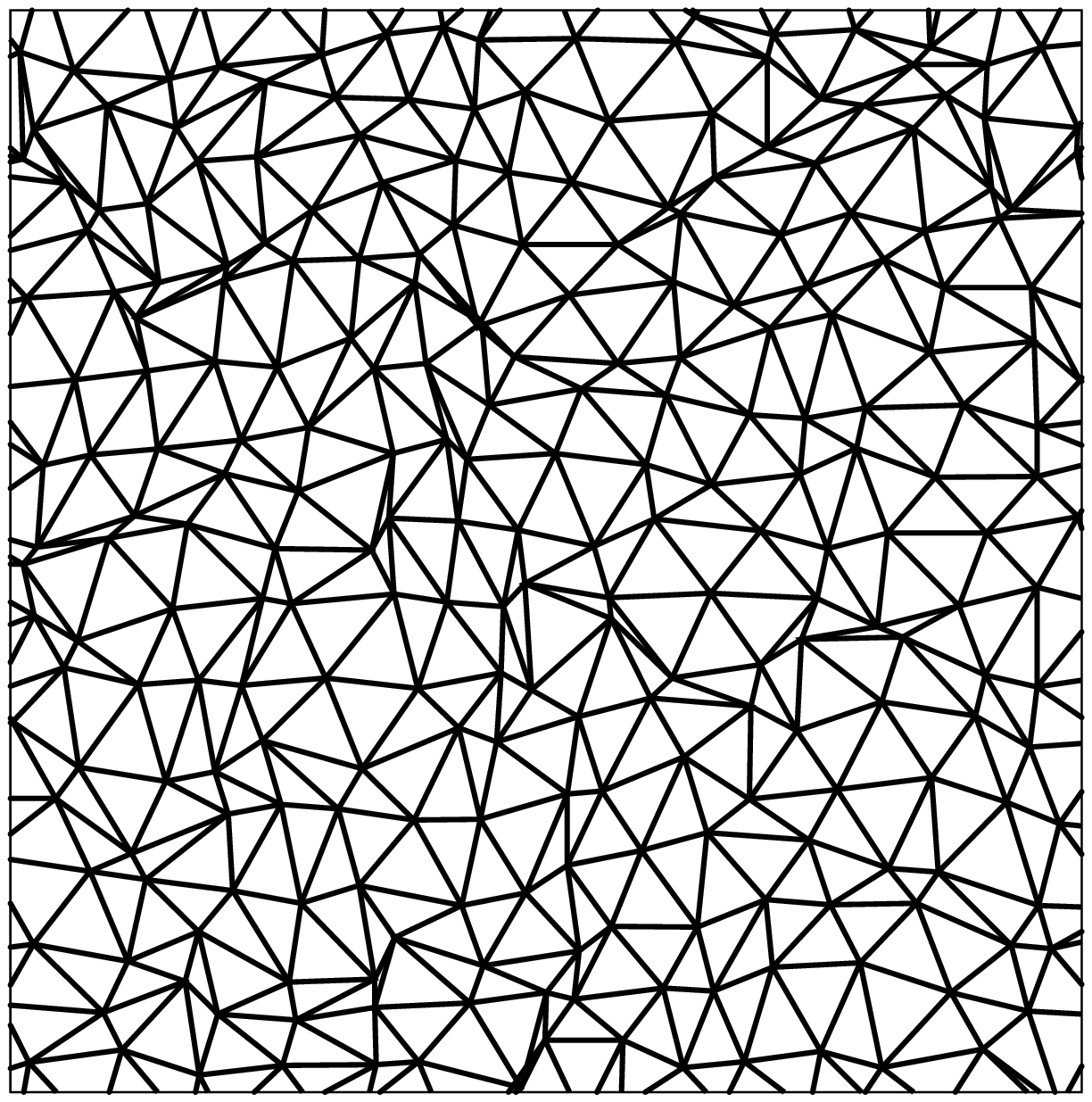}}}
%\vspace{-1.0cm}
\centerline{\large (b)}
\vspace{0.5cm}
\caption {\protect Configurations corresponding to different values of
  the ratio $\gamma$ between the maximal and actual extensions of the
  net: (a) $\gamma=1.1$, (b) $\gamma=1.5$. Only part of the net is
  shown in the figures.}  
\label{picture1}
\end{figure}
Our simulations were performed on systems consisting of 1600 atoms
which were bonded to form a triangular 2D net. The topology of the net
is such that the mean positions of the atoms form a regular triangular
lattice with lattice spacing $b_0$, while each pair of nearest
neighbor atoms is connected by a tether whose maximal extension is
$b\geq b_0$. Periodic boundary conditions, which fixed the volume and
prevented the net from collapsing, were applied. We denote by
$\gamma\equiv b/b_0$, the ratio between between the maximal and actual
extensions of the net. Typical equilibrium configurations
corresponding to two values of $\gamma$ are depicted in
Fig.~\ref{picture1}. We generated the MC configurations using a new
updating scheme, recently proposed by Jaster \cite{jaster}, in which
the conventional Metropolis step of a single particle is replaced by a
collective step of chain of particles. At each MC time unit we made
1600 move attempts (with acceptance probability $\sim 0.7$), where at
each attempt a new atom was selected randomly. (On the average, each
atom was chosen once in a MC time unit.) Correlations between
subsequent configurations were estimated from the autocorrelation
function of the amplitude of the longest-wavelength phonon in the
systems (both longitudinal and transverse phonons were checked). For
all $\gamma$ values, we found that after less than 1000 MC time units,
the memory of the initial configuration is completely lost. We
measured the stress and elastic constants for many values of
$\gamma$. For each $\gamma$, we averaged the relevant quantities over
a set of $1.5\times10^7$ configurations separated from each other by 3
MC time units. We also evaluated the standard deviations of the
averages. The error bars appearing in the graphs which present our
results correspond to one standard deviation. More technical details
of the simulations, as well as a detailed explanation on the formalism
used in this work, were given in another publication \cite{fk}.  

\begin{figure}[htb]
\epsfysize=18.5\baselineskip
\centerline{\hbox{
      \epsffile{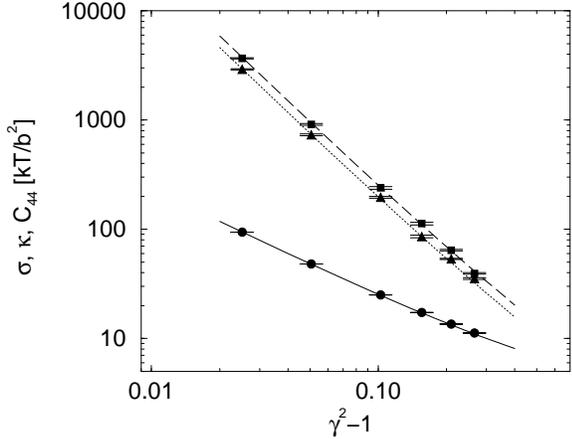}  }}
\caption {\protect Numerical results for the stress $\sigma$
  (circles), the bulk modulus $\kappa$ (squares), and the elastic
  constant $C_{44}$ (triangles), as a function of the ratio $\gamma$
  between the maximal and actual extensions of the net. Results are
  in $kT/b^2$ units. The solid, dashed and dotted curves depict the
  expressions on the right sides of
  Eqs.~(\protect\ref{presrel})--(\protect\ref{murel}), respectively
  [with $A=1.80$ in Eq.~(\protect\ref{murel})].}  
\label{fisherg1}
\end{figure}

When the net is fully extended ($\gamma=1$), atoms cannot leave their
mean lattice positions. Entropy, therefore, vanishes,
while the stress and elastic constants diverge. For slightly larger
values of $\gamma$, atoms are restricted to small thermal fluctuations
around their lattice positions, as in Fig.~\ref{picture1}~(a). A
similar atomic-level picture appears in hard disks (2D ``hard
spheres'') solids for densities proximal to the close-packing
density. In fact, the FN and the hard disks (HD) problems are closely
related: In the latter (HD) the centers of the disks are not allowed
to approach their neighbors a distance smaller than $a$, the diameter
of the disks, while in the former (FN) atoms are not allowed to depart
from their neighbors a distance larger the maximal extension of the
bond, $b$. For HD solids, one can define the ratio $\delta=a/b_0\leq
1$ between the diameter of the disks, $a$, and the mean lattice
separation, $b_0$. In the limits $\gamma\rightarrow 1$ and
$\delta\rightarrow 1$ (corresponding to the FN and HD problems,
respectively), the elastic constants of both systems coincide, as can
be seen from the following argument: Let $\Pi_{\rm FN}$ and
$\Pi_{\rm HD}$ be phase spaces of allowed configurations of a FN with
a certain value of $\gamma$ and of a HD solid with $\delta=1/\gamma$,
respectively. Each configuration in one of these phase spaces can
be described by the set $\{{\bf u}_i\}$ of deviations of either the
atoms of the net or the centers of the disks from their mean lattice
positions. In the $\gamma$, $\delta\ \sim 1$ asymptotic regimes, we
can assume that the size of all the deviations is much smaller than
the lattice spacing, $b_0$. One can easily check that if the set
$\{{\bf u}_i\}$ represents an allowed microscopic configuration of the
FN, then the set $\{-{\bf u}_i\}$ almost always corresponds to an
allowed configuration of the HD system. Moreover, by this
transformation we can generate almost all the configurations of
$\Pi_{\rm HD}$. The measure of the subgroup of configurations for which
the mapping $\{{\bf u}_i\}\longleftrightarrow \{-{\bf u}_i\}$ between
the two problems does not apply, diminishes proportionally to $\langle
{u_i}^2\rangle/{b_0}^2$. Thus, the mapping $\{{\bf
  u}_i\}\longleftrightarrow \{-{\bf u}_i\}$ is asymptotically a {\em
  one-to-one}\/ transformation from $\Pi_{\rm FN}$ onto $\Pi_{\rm
  HD}$. Since for both systems the Helmholtz free energy $F$ is equal
to $-kT\ln|\Pi|$, where $|\Pi|$ is the volume of the $2N$-dimensional
configuration phase space ($N$ is the number of atoms), and since the
Jacobian of the above transformation is unity, we readily find that
the free energies $F_{\rm HD}$ and $F_{\rm FN}$ of the HD and FN
systems, respectively, are related by   
\begin{eqnarray}
\nonumber
F_{\rm FN}(N,\gamma)\simeq F_{\rm HD}(N,\delta=1/\gamma),\ \ \ \ \
{\rm for\ } \gamma\sim 1. 
\end{eqnarray}
Suppose now that both systems are slightly deformed from their
reference states. The displacements of the atoms from their mean
lattice positions can be divided into the set $\{{\bf u}_i\}$ of
thermal fluctuations and the set $\{{\bf v}_i\}$ of small changes in
mean lattice positions caused by the deformation. The transformation
between $\Pi_{\rm FN}$ and $\Pi_{\rm HD}$, in this case, maps both
$\{{\bf u}_i\}$ to $\{-{\bf u}_i\}$ and $\{{\bf v}_i\}$ to $\{-{\bf
  v}_i\}$. The $\{{\bf v}_i\}$ mapping is equivalent to the reversal
of the strain applied on the system. We therefore find that $F_{\rm
  HD}$ and $F_{\rm FN}$ will be equally modified, provided that
opposite strains are applied on the FN and HD systems. The following
asymptotic relations between the stress and elastic constants of these
systems follow immediately: 
$\sigma_{\rm FN}(\gamma)\simeq P_{\rm HD}(1/\gamma)$, $\kappa_{\rm
  FN}(\gamma)\simeq \kappa_{\rm HD}(1/\gamma)$ and ${C_{\rm
    FN}}_{44}(\gamma)\simeq {C_{\rm HD}}_{44}(1/\gamma)$.   
These relations are very useful since the asymptotic expressions for
$P_{\rm HD}$, $\kappa_{\rm HD}$ and ${C_{\rm HD}}_{44}$ are available
\cite{stisal}, and can be used to find the stress and bulk modulus of
the FN. This gives us the {\em exact}\/ expressions 
\begin{eqnarray}
\label{presrel}
\sigma_{\rm FN}(\gamma) &\simeq & 
\frac{4/\sqrt{3}}{(\gamma^2-1)}\frac{kT}{b_0^2},\\  
\label{kapparel}
\kappa_{\rm FN}(\gamma) &\simeq &
\frac{4/\sqrt{3}}{(\gamma^2-1)^2}\frac{kT}{b_0^2}.
\end{eqnarray}
For the elastic constant $C_{44}$, Ref.\cite{stisal} finds only the
asymptotic functional form, and therefore for our problem we have 
\begin{equation}
\label{murel}
{C_{\rm FN}}_{44}(\gamma)\simeq
\frac{A}{(\gamma^2-1)^2}\frac{kT}{b_0^2},
\end{equation}
with an unknown constant $A$. Our numerical results, presented in
Fig.~\ref{fisherg1}, confirm these relations, which seem to be
accurate over quite a large range of $\gamma$ values. In
Eq.~(\ref{murel}), we use the value $A=1.80\pm0.02$ obtained by
fitting the asymptotic expression for $C_{44}$ to the three data
points corresponding to the smallest $\gamma$ values. Note that
while in Eqs.~(\ref{presrel})--(\ref{murel}), $P_{\rm HD}$,
$\kappa_{\rm HD}$ and ${C_{\rm HD}}_{44}$ are expressed in units of
$kT/b_0^2$, in Fig.~\ref{fisherg1} they are given in units of
$kT/b^2$. In this representation, the stress and elastic constants of
the FN are scaled to depend on the parameter $\gamma$ alone. 
\begin{figure}[htb]
\epsfysize=17.5\baselineskip
\centerline{\hbox{
      \epsffile{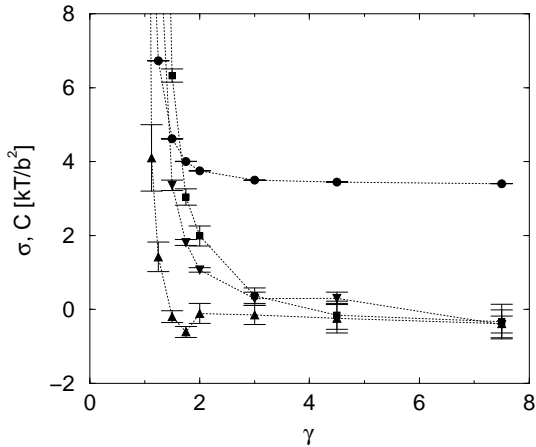}  }}
\caption {\protect Numerical results for the stress $\sigma$
  (circles), and the elastic constants $C_{11}$ (squares), $C_{12}$
  (triangles pointing up) and $C_{44}$ (triangles pointing down), as a
  function of the ratio $\gamma$ between the maximal and actual
  extensions of the net. Results are in $kT/b^2$ units. The lines are
  guides to the eye.}  
\label{fisherg2}
\end{figure}
Fig.~\ref{fisherg2}, shows the dependence of the stress and elastic
constants on $\gamma$ for weakly stretched nets. We observe a
spectacular decay of elastic constants to almost zero for $\gamma\sim
3$, and at the same time we note that the stress becomes independent
of $\gamma$. The very fact of decrease of elastic constants with
increasing $\gamma$ should not be surprising, because it is
intuitively clear that larger $\gamma$ represent a more ``loose'' and
more ``weak'' solid. However, almost vanishing values already at
$\gamma\sim3$ are {\em not} direct consequences of the ``weakness'' of
the solid, but of the fact that a ``loose'' network can be
approximated by a network of {\em Gaussian springs}. Gaussian spring
is a linear spring of vanishing unstressed length. The energy of such
a spring, $E=\frac{1}{2}Kr^2$, is simply proportional to its squared
end-to-end distance, $r^2$. We will show that elastic solid formed by
such springs has {\em exactly}\/ vanishing elastic constants,
independently of the value of the spring constant $K$. Thus, the
effect observed in Fig.~\ref{fisherg2} is an indication of the
Gaussian nature of the system.

It is easy to calculate the elastic properties of Gaussian networks at
$T=0$: The stresses of such networks depend on their topologies,
namely on the details of the connectivity between the atoms and on the
values of the springs constants between them. For 2D networks the
stresses are not modified due to homogeneous changes in the size of
the net, since the force exerted on the surface grows (diminishes)
linearly with the length of the boundaries. Moreover, at $T=0$, the
free energy, $F$, coincides with the internal energy  $E=\sum_{{\rm
    bonds}\, \langle\alpha\beta\rangle}\frac{1}{2}K_{\alpha\beta}
(r^{\alpha\beta})^2$, where $r^{\alpha\beta}$ is the length of the
bond connecting atoms  ``$\alpha$'' and ``$\beta$'', and 
$K_{\alpha\beta}$ is the spring constant assigned to this bond. From
Eq.~(\ref{trans}) it is obvious that the energy expansion in the
strain variables includes only linear terms in $\eta$, and hence, by
comparing with Eq.~(\ref{expan}), $C_{ijkl}(T=0)\equiv 0$. This
identity, as well as the size independence of the stresses, hold at
any other temperature since Gaussian networks have the interesting
feature that their stress and elastic constants are temperature
independent! For the stresses this feature is readily understood: The
stresses can be expressed as the averages of quantities which are
linear in the coordinates of the atoms. When the statistical weights
of the distribution are Gaussian, i.e., an exponent of a quadratic
form of the coordinates, these averages coincide with the most
probable values, namely their values at equilibrium. The temperature
independence of the elastic constants then follows immediately, since
the latter are just the derivatives of the stress components. 

\begin{figure}[htb]
\epsfysize=18.5\baselineskip
\centerline{\hbox{
      \epsffile{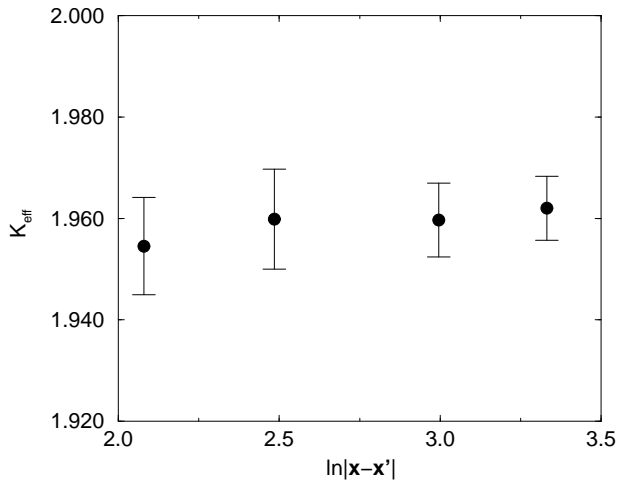}  }}
\caption {\protect The effective spring constant $K_{\rm eff}$,
  extracted from MC measurements of $r_{{\bf xx'}}^2=\langle|{\bf
    r}({\bf x})-{\bf r}({\bf x'})|^2\rangle$ [see
  Eq.~(\protect\ref{kef})]. The error bars correspond to one standard
  deviation in the estimated value of $r_{{\bf xx'}}^2$.} 
\label{radius}
\end{figure}

The similarity between non-stressed tethered and Gaussian
one-dimensional (1D) nets, i.e., linear polymers, is a consequence of
central limit theorem \cite{weiner}. For topologically two-dimensional
regular (non-random) nets, such similarity was demonstrated by Kantor
et al.~\cite{kkn}: In both tethered and Gaussian two-dimensional
nets, the mean squared distance in the embedding space, 
$r_{{\bf xx'}}^2=\langle|{\bf r}({\bf x})-{\bf r}({\bf x'})|^2\rangle$,
between two distant points whose internal positions in the net
(measured in lattice constants) are ${\bf x}$ and ${\bf x'}$, grows
proportionally to $\ln|{\bf x}-{\bf x'}|$. One can define the
effective spring constant, $K_{\rm eff}$, as the value of $K$ of a
Gaussian network with the same connectivity and statistical
properties as of the tethered network. The value of $K_{\rm eff}$ is
extracted from the ratio of the mean squared distance, $r_{{\bf
    xx'}}^2$, between two points ${\bf x}$ and ${\bf x'}$ on the FN,
and the mean squared distance ${\tilde{r}_{{\bf xx'}}}^2$ between the
same two points on a Gaussian network of unit spring constants: 
\begin{equation}
K_{\rm  eff}={\tilde{r}_{{\bf xx'}}}^2/r_{{\bf xx'}}^2. 
\label{kef}
\end{equation}
${\tilde{r}_{{\bf xx'}}}^2$ can be calculated exactly, while the value
of the corresponding $r_{{\bf xx'}}^2=\langle|{\bf r}({\bf x})-{\bf
  r}({\bf x'})|^2\rangle$  can be extracted from MC simulations of the
FN with free boundaries conditions (i.e., in the absence of external
pressure). We simulated a FN of $56^2=3136$ atoms and measured (using
$10^7$ different configurations) $r_{{\bf xx'}}^2$ for several pairs
of points ${\bf x}$ and ${\bf x'}$ at different lattice
separations. With these measurements we evaluated the effective spring
constants [using Eq.~(\ref{kef})], and found, as shown in
Fig.~\ref{radius}, that for the FN model $K_{\rm eff}\simeq 1.96\
kT/b^2$. In order to support our conclusion about the crossover into
the Gaussian regime, we need to show that the constant value to which
the stress drops in Fig.~\ref{fisherg2}, is just the stress applied by
a Gaussian net with spring constants $K_{\rm eff}$ calculated for {\em
  non-stressed}\/ FN. For a Gaussian net with $K\simeq 1.96$, one
finds that $\sigma=\sqrt{3}K\sim 3.39\ kT/b^2$ which indeed coincides
with the value of $3.4\ kT/b^2$, extracted from Fig.~\ref{fisherg2}.  
The persistence of the Gaussian regime to intermediate values of
$\gamma$ ($\gamma\sim 3$), is not unique for 2D nets. Such behavior is
also found, for instance, in 1D polymers. Let us consider, for a
moment, a chain of $N\gg 1$ tethers of maximal length $b$, which is
stretched by a force $f$, to an end-to-end length $l=Nb_0$. It is a
well known fact that this chain will be Gaussian, i.e., $f$ and $l$
will be proportional to each other, provided that $l$ does not exceed
the order of magnitude of the root mean square size of the chain:
\begin{equation}
\label{1dcrit} 
l=Nb_0\lesssim\sqrt{N}b.
\end{equation}
Yet, one must understand that in order to observe Gaussian elastic
behavior, it is not essential to apply this criterion (\ref{1dcrit})
to the whole chain, but only to small segments of it. If there exist a
certain length scale at which the potential between the atoms becomes
effectively quadratic, i.e., can be replaced by a Gaussian spring,
then the whole chain is like a chain of Gaussian springs, and
therefore it is itself Gaussian. For a linear polymer chain, the
effective potential between non-neighboring atoms is calculated by
integrating out the spatial degrees of freedom of the atoms located
between them. Such calculations are usually done iteratively, where on
each ``rescaling'' step every second atoms is integrated out. It
appears that even elementary potentials which are very different from
parabola, are brought into a parabolic form within a few ``rescaling''
steps. For the specific potential used in this work, three steps are
sufficient, which means that a segment of $N\sim 10$ tethers may be
justly considered as an effective Gaussian spring. Similarly to a
macroscopically large chain, we expect that the Gaussian nature of
this segment will persist as long as it is stretched to a length which
does not exceed its root mean square size, namely, as long as
$10b_0\lesssim \sqrt{10}b$ [see criterion (\ref{1dcrit})]. This
relation gives the lower limit, $\gamma=b/b_0\gtrsim
\sqrt{10}\sim3$, of the Gaussian regime of a 1D chain of tethers. 
For a 2D regular phantom net, the effective potential becomes
approximately parabolic also for a distance of number of bonds, $N\sim
10$ \cite{kkn}. Root mean square distance between two such points is
$b\sqrt{\ln N}$. Thus, in order to observe Gaussian elastic behavior,
we require that $10b_0\lesssim b\sqrt{\ln 10}$, or,
$\gamma=b/b_0\gtrsim 10/\sqrt{\ln 10}\sim 4$, which is consistent
with the value $\gamma\sim 3$, observed in Fig.~\ref{fisherg2}.      

In summary, we have applied a new ``fluctuation'' formalism to MC
determination of the stress and elastic constants of stretched
tethered networks. These systems provide a convenient framework for
studying the entropic contribution to elasticity in real polymeric
systems. The Gaussian nature of entropic elasticity, observed for
non-stressed phantom nets, was also found when stress was applied. It
breaks only for highly extended networks, close to their 
full-extension. This point has interesting implications to the problem
of the critical elastic behavior of gels right above the gel-point. As
already mentioned at the first paragraph of this paper, recently it
was suggested by Plischke, Jo\'os and co-workers that this behavior is
dominated by entropy \cite{plis}. These authors studied numerically
(using a different technique) the elastic behavior, at $T\neq 0$, of
bond diluted (percolating) systems at which only a fraction $p$ of the
bonds were present. Their results in $2D$ for the critical exponent
$f$ characterizing the growth of the shear modulus above the
percolation threshold $p_c$, $\mu\sim(p-p_c)^f$, match (within the
range of  error), the known result for the exponent $t$ describing the
conductivity of random resistors network $\Sigma\sim(p-p_c)^t$. The
question is whether this result is universal. For Gaussian networks
the identity, $f=t$, can be proven rigorously \cite{kandegen}. One can
further argue that this result also applies to other types of
interactions, provided that above a certain finite length-scale, the
network is {\em effectively}\/ Gaussian. We have shown here that this
property is not always insured. In a percolation problem, the elastic
backbone is inhomogeneous and includes very tenuous parts where the
tension applied to the network is distributed between very few
strands. Such strands may deviate from Gaussian behavior when high
stress is applied. Further complications can arise from excluded
volume effects which have not been discussed here at all.

This work was supported by the Israel Science Foundation through Grant
No. 177/99.  
%-----------------------------------------------------------
% References
%-----------------------------------------------------------

\end{multicols}

\end{document}